\documentclass[conference]{IEEEtran}
\IEEEoverridecommandlockouts
\usepackage{cite}
\usepackage{amsmath,amssymb,amsfonts}
\usepackage{algorithmic}
\usepackage{graphicx}
\usepackage{subcaption}
\usepackage{textcomp}
\usepackage{xcolor}
\graphicspath{{images/}}
\def\BibTeX{{\rm B\kern-.05em{\sc i\kern-.025em b}\kern-.08em
    T\kern-.1667em\lower.7ex\hbox{E}\kern-.125emX}}
\begin{document}

\title{Synchrophasor Data Anomaly Detection on Grid Edge by 5G Communication and Adjacent Compute\\
\thanks{This work is part of 5G Energy FRAME project funded by the U.S. Department of Energy (DOE) Office of Science through the Advanced Scientific Computing Research (ASCR) Program. PNNL is a multi-program  national laboratory operated for the U.S. Department of Energy (DOE) by Battelle Memorial Institute under contract no. DE-AC05-76RL0-1830.}
}

\author{\IEEEauthorblockN{Chuan Qin \IEEEauthorrefmark{1}, Dexin Wang \IEEEauthorrefmark{1}, Kishan Prudhvi Guddanti \IEEEauthorrefmark{1}, Xiaoyuan Fan\IEEEauthorrefmark{1}, Zhangshuan Hou\IEEEauthorrefmark{2}}
\IEEEauthorblockA{\IEEEauthorrefmark{1}\textit{Electricity Infrastructure \& Building Division}, \IEEEauthorrefmark{2}\textit{Earth Systems Science Division}}
\textit{Pacific Northwest National Laboratory}, Richland, USA \\
emails: {\{chuan.qin$|$dexin.wang$|$kishan.g$|$xiaoyuan.fan$|$zhangshuan.hou\}}@pnnl.gov}

\maketitle

\begin{abstract}
The fifth-generation mobile communication (5G) technology offers opportunities to enhance the real-time monitoring of grids. The 5G-enabled phasor measurement units (PMUs) feature flexible positioning and cost-effective long-term maintenance without the constraints of fixing wires. This paper is the first to demonstrate the applicability of 5G in PMU communication, and the experiment was carried out at Verizon non-standalone test-bed at Pacific Northwest National Laboratory (PNNL) Advanced Wireless Communication lab. The performance of the 5G-enabled PMU communication setup is reviewed and discussed in this paper, and a generalized dynamic linear model (GDLM) based real-time synchrophasor data anomaly detection use-case is presented. Last but not least, the practicability of implementing 5G for wide-area protection strategies is explored and discussed by analyzing the experimental results.
\end{abstract}


\section{Introduction}
\IEEEPARstart{A}{dvanced} telecommunication technology enables flexible and cost-effective communications between sensors, intelligent electronic devices (IEDs), substation automation systems (SASs), and merging units (MUs) through wireless broadband access over electrical power system infrastructures. Phasor measurement unit (PMU) as one type of IED has been widely deployed in power system state monitoring and analysis. It also can be leveraged in transactive distribution systems for investigating the bidirectional power flow caused by renewable energy integration \cite{5535240, 5524055, 9905351}. The deployment of PMUs and micro-PMUs in the sprawling and radial electrical grids facilitates digital intellectualization in the applications of monitoring, analysis, control, and protection \cite{9917615, 9787983}. Compliance with the requirements of IEC 61850 \cite{IEC61850}, an energy-efficient telecom power can protect against grid power interruptions and fluctuations and help operators reduce operational expenditure (OpEx) and their carbon footprint. Utilizing grid-edge sensor data communication and delay-tolerant power system applications, such like fine-grained transactive control of distributed energy resources (DERs), is another advantageous aspect. Enabling wireless communication in PMU data streaming can extend the coverage range and enhance the flexibility in selecting locations \cite{6083199}.


The up-to-date PMU configurations involve cumbersome wired connections, incurring substantial costs and lacking flexibility. In contrast, the broadband cellular network capabilities of the advent of fifth-generation (5G) technology, with its ultra-reliable low latency communication (URLLC) features, hold promise in effectively addressing these issues. The commissioning wireless communication network can cluster PMUs based on the system topology to effectively support data exchange \cite{6469217}. Meanwhile, 5G can ensure end-to-end communication security \cite{9797119}. This technology can be aptly harnessed to facilitate PMU communication and resolve the aforementioned challenge. 


This paper introduces an operational framework for 5G-enabled PMU data communication alongside a PMU streaming data anomaly detection application relying on 5G network edge computation. The contribution of this research is underscored by its initial experimental validation of 5G implementation within PMU communication. The advantages and drawbacks of integrating 5G wireless technology into PMU applications are thoroughly explored through experimentation and analysis. Moreover, the article presents a practical case of machine learning (ML) based anomaly detection in PMU data using 5G edge computation. The collected findings undergone thorough analysis.

\section{PMU Communication Overview}
The PMU conducts measurements encompassing current, voltage, and frequency to estimate phasors within an electrical system. This is facilitated by integrating the global positioning system (GPS), as stipulated by IEEE Std. 1344 \cite{943067}. Compared to supervisory control and data acquisition (SCADA) systems, PMUs exhibit a notable high-speed data rate capable of capturing information at several hundred frames per second. This exceptional capability empowers the monitoring and managing of the stability and dependability of electrical power systems.
\begin{figure*}[!ht]
    \centering
    \includegraphics[width=\linewidth]{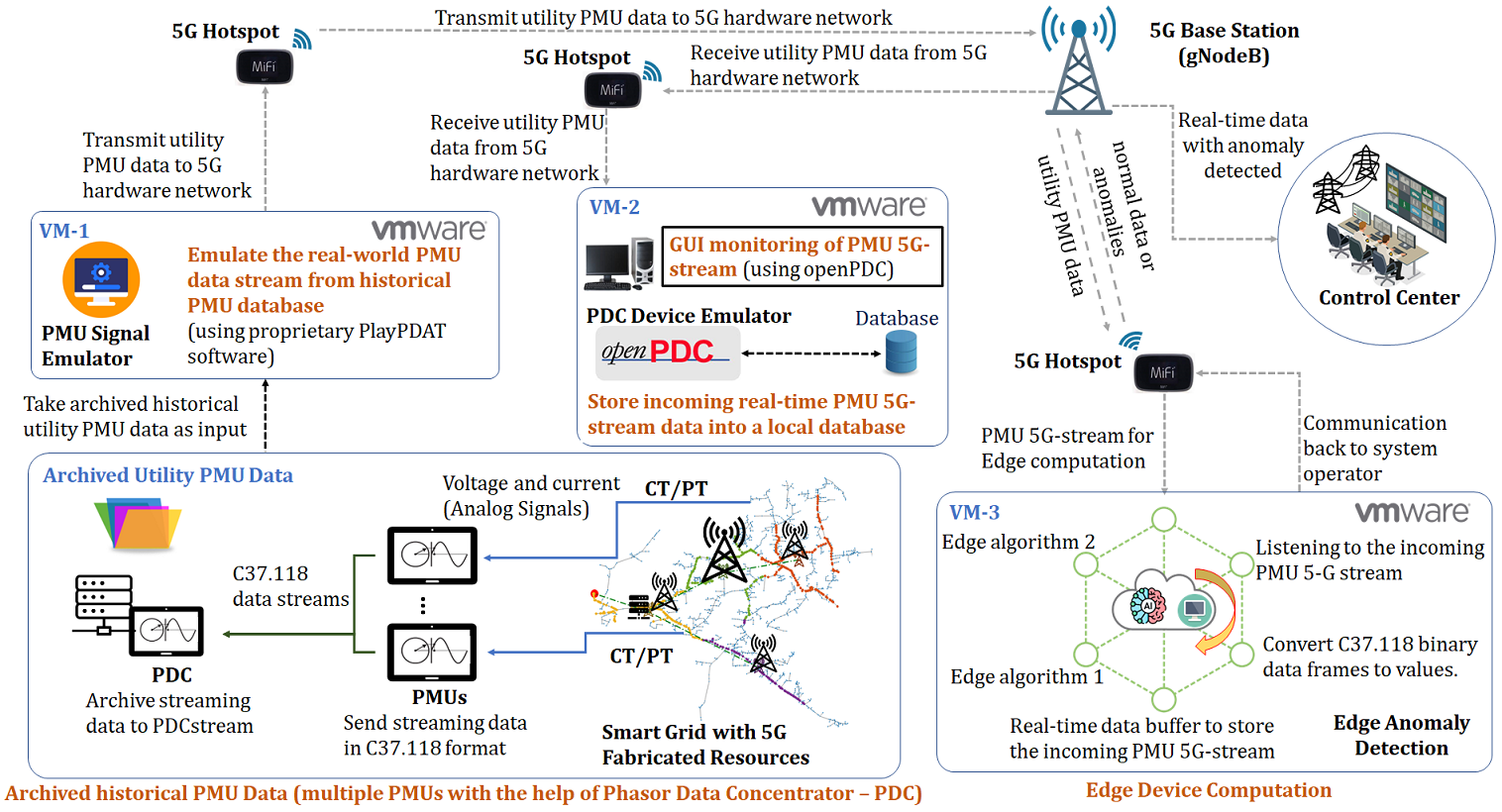}
    \caption{The configuration of 5G infrastructure for PMU communication with edge computation.}
    \label{fig:overview_5g_diagram}
\end{figure*}

The PMU data conforms to the specifications outlined in IEEE Std. C37.118.2-2011 for Synchrophasor Data Transfer for Power Systems \cite{ieeestd}. This standard succinctly defines distinct message types, including data, configuration, header, and command, for real-time synchrophasor communication in clause 6. Moreover, the standard specifies that messaging can be employed with any suitable communication protocol for real-time interactions among PMUs, phasor data concentrators (PDCs), and other applications. Notably, the standard refrains from enforcing any limitations on communication systems or media. Nevertheless, according to Annex C, the communication system should consider the following aspects: C.1) Communication bandwidth and C.2) communication delay.

C37.118.1a-2014 \cite{6804630} Annex C.1 meticulously delineates that the frame size fluctuates based on the number of phasors and analog and digital words encompassed within the frame. Frame sizes typically vary from 40–70 bytes for frames from a single PMU, expanding to over 1000 bytes for frames from a PDC incorporating data from numerous PMUs. C37.118 Annex C.2 provides the requirements for PMU communication delay, and it defines that the communication delay begins by extracting bits from the PMU. In instances of RS-232 serial communication, the delay can dwell in milliseconds, yet at network velocities, it assumes insignificance. As an illustration, data transmitted over 500 miles would encounter a mere 3 ms to 5 ms delay—relatively minor figures that are foreseeable based on the characteristic of the communication network. The minimum anticipated overall delay stands at approximately 20 ms (excluding supplementary filtering). In real-time systems, the maximum delay could extend into the order of 10 seconds or beyond. A standard system, spanning from PMU to PDC, typically falls within the 20 ms to 50 ms range. Each subsequent level introduces incremental PDC processing and waiting periods, likely ranging from 30 ms to 80 ms (in addition to the PMU to PDC time).

\section{5G-Enabled PMU Communication Framework}
This section discusses the configuration of the lab testing environment for 5G-enabled PMU communication. The test consists of a PMU streaming data generator, a PMU data receiver, and an edge application server. The emulators and software are deployed in three separate virtual machines (VMs) that operates on ESXI VM host. Fig. \ref{fig:overview_5g_diagram} shows the combination of hardware and software environment. The Std. C37.118 stipulates that messaging is compatible with any appropriate communication protocol to facilitate real-time interaction among PMUs, PDCs, and other applications \cite{ieeestd}. In this context, the user datagram protocol (UDP) broadcast communication protocol has been utilized to transmit PMU data with time-sensitive communication over a 5G network. 


\noindent\underline{\textbf{Software Setup}}\vspace{.2em}: To test the functionality of 5G in PMU data communication, 4 software are installed in VMs -- PlayPDAT, OpenPDC \cite{openpdc}, Wireshark \cite{wireshark}, and Python. PlayPDAT is a LabVIEW-based executable software that emulates the PMU data generation based on historical utility data and transmits the C37.118 streaming PMU data to a designated destination with the specified IP address and port via user datagram protocol UDP. It is installed on \textbf{VM-1} as the signal generator. The open-source openPDC is deployed on \textbf{VM-2} to receive the data transmitted over a 5G network. The Wireshark is employed and used for analyzing the packets. On \textbf{VM-3}, a Python-developed streaming data interface is deployed to listen to the streaming PMU data from 5G. For data processing, anomaly detection, and early awareness, an ML-based generalized dynamic linear model (GDLM) has been created and incorporated.

\noindent\underline{\textbf{Operating System Setup}}\vspace{.2em}: Two operating systems (OS) were installed for this experiment -- Microsoft Windows 10 (64-bit) and Ubuntu Linux (64-bit). PlayPDAT and openPDC are executed on Windows 10 OS VM, and Linux acts on an edge server for running the ML programs.

\noindent\underline{\textbf{Network Setup}}\vspace{.2em}: The three VMs PMU emulator (\textbf{VM-1}), PDC emulator (\textbf{VM-2}), and GDLM anomaly detection tool (\textbf{VM-3}) are operating on the ESXI host. The VM clocks are synchronized by precision time protocol (PTP), also known as IEEE 1588 \cite{4579760}. PTP has highly accurate time synchronization that is usually applied in industrial automation and telecommunications. The essential operation of PTP involves exchanging messages between a master clock and a set of slave clocks. The master clock sends synchronization messages to the slave clocks, which use the information to adjust their clocks to match the master clock. PTP can achieve synchronization accuracy in the sub-microsecond range, which is sufficiently precise for latency measurement compared to the PMU reporting rate. Moreover, each VM is connected to different virtual network adapters with separate virtual local area networks (VLAN). This ensures that the multiple VMs/nodes from different LANs are configured to communicate through the unique logical 5G network. 
\section{5G Based Edge Computation Use Case:  Online Anomaly Detection}
This section presents the deployment of an anomaly detection algorithm on a 5G network test bed and its methodology. As shown in Fig.~\ref{fig:overview_5g_diagram}, the utility PMU data received by the edge computing node at \textbf{VM-3} is streaming type data in binary format. The detected anomalies are sent to the control center through a 5G network that helps system operators to classify the events and make the operation decisions. This work demonstrates an ML application on the edge computing node located at \textbf{VM-3}. More specifically, we present the anomaly event detection use case using real-world data.

\subsection{Emulation of Real-world Anomalous Event with 5G-based Hardware Network} 
\label{subsection:emulation_layer_for_pmu_measurements}
In order to showcase the anomaly detection use case using a 5G-based hardware network and real-world data, we need the data received by the anomaly detection computation node (\textbf{VM-3}) to contain anomalous events from the real-world. We successfully emulated this behavior using the proposed testbed. This subsection presents the approach used for this emulation and the dataset information.

\subsubsection{\textbf{Data}}
\label{subsubsec:data}
The data utilized in this work contains the frequency attribute of one utility's PMU signals that belong to Western Interconnection. There are four attributes, including the voltage, angle variation, frequency, and rate of change of frequency (ROCOF) at each PMU. The utility PMU data is stored in PDAT format \cite{faris2016bpa} based on IEEE Std. C37.118.2-2011 data frames \cite{ieeestd}. This storage in PDATA format is done when the real-world event took place (as shown in the ``Archived historical PMU data" block in Fig.~\ref{fig:overview_5g_diagram}). This anomalous event is validated against the frequency event database maintained by North American Electric Reliability Corporation (NERC) Resource Subcommittee. Specifically, the PMU data for the hour in which the event occurred is selected to demonstrate the proposed anomaly detection edge device identifying the event correctly while using a 5G-based hardware network.

\subsubsection{\textbf{Emulation of 5G Communication Network During Anomalous Event Using Historical Data and Proposed Test Bed Framework}} 
\label{subsubsec:emulate_comm}
Section~\ref{subsubsec:data} discusses the historical data with the anomalous event. However, for real-time anomaly detection demonstration using 5G, we need to emulate the event as well as the entire communication network (with 5G capability). To achieve this, as shown in Fig.~\ref{fig:overview_5g_diagram}, we used playPDAT software to read the local archived historical data on \textbf{VM-1} and transmit this historical data to the 5G server as real-time PMU measurements in binary data frames (IEEE Std. C37.118.2-2011 \cite{ieeestd}).

\subsection{Detection of Emulated Real-world Anomalous Event with 5G-based Hardware Network} The detection algorithm is implemented on edge device (\textbf{VM-3}) as shown in Fig.~\ref{fig:overview_5g_diagram}. This detection algorithm is an online process, i.e., the real-time measurements are being streamed from the 5G server to the edge device. The edge device must process/identify the anomalies while polling its response to the 5G server. We first implemented a listening module that listens to the incoming 5G data stream to facilitate this online process. This 5G data stream contains data frames/messages in IEEE Std. C37.118 binary format. Second, we implemented a decoding module to convert the binary data frames into numerical values for the frequency attribute of the PMU measurements. Third, the converted measurement data is saved into a local data buffer to create sliding windows of measurement data (used for anomaly detection algorithm as input).

\subsubsection{\textbf{Fast Model Fitting Using GDLMs}}
\label{subsubsec:fitting}
The object of GDLM is, at any given time $t$, given some data $D_t$, a model is learned for $\left(\theta_t\ |\ D_t\right)$. The system model is then updated considering the current time $t$ i.e., $\left(\theta_{t+1}\ |\ \theta_{t}\right)$ and next observation is estimated using $\left(y_{t+1}\ |\ \theta_{t+1}\right)$. More specifically, GDLM uses the state space model equations shown below. It contains an observation equation ($y_t$) and a model equation ($\theta_t$).
\begin{align}
    y_t &= \textbf{H}_t \cdot \theta_t + \epsilon_t\ \sim N(0, R_t) \\
    \theta_t &= \textbf{M}_t \cdot \theta_{t-1} + E_t\ \sim N(0, Q_t)
\end{align}
where $y_t$ is the observation vector whose length equals the sliding window for time series analysis. $\theta_t$ is an unobserved state vector of the system that can change over time (different sliding windows). $\textbf{M}_t$ is the linear system operator that dictates the changes in state vector at time $t$. $\textbf{H}_t$ is the observation matrix that helps transform a given state vector into an observation vector. $R_t$ and $Q_t$ are the covariance matrices of the Gaussian distribution with zero means representing the errors in observation and model equations.

In this work, we modeled the GDLM as a second-order polynomial function. The second order GDLM is modeled as follows where $\textbf{H}_t$ is given by $(1,\ 0)$ and $\textbf{M}_t$ is given by $\begin{bmatrix}1 & 1 \\ 0 & 1\end{bmatrix}$. The prior for the mean is set as the median of the observation vector $y_t$, and the prior for covariance as identify matrix. GDLMs are solved using the following steps, 1) Kalman Recursion forward step for model fitting (estimation of different conditional probability distributions), 2) Kalman Smoothing backward step. 

The proposed work requires the anomaly detection method to be fast given that it involves utilizing real-world measurements in a streaming manner. The GDLM used in this work is fast due to its closed-form analytic solution nature, as explained in \cite{ren2018online}. The model is implemented using the pyDLM python package, which is optimized for fast model fitting and inference. 
The pyDLM package has fast model fitting (which is the time-consuming step in GDLMs) because, different from usual Kalman filters, it uses a modified Kalman filter technique that does not require tuning of two parameters 1) error covariance matrix and 2) observational matrix, making it an efficient computation. 
This is done by using the discounting factor implementation. We used $0.95$ as the discounting factor as the objective of anomaly detection revolves around real-world measurement data and relies heavily on observations. This quickly helped identify the anomalous event using the proposed 5G test bed.

\subsubsection{\textbf{Online Prediction/Anomaly Detection}}
\label{subsubsec:prediction}
As described in Section~\ref{subsubsec:fitting}, the Kalman filter forward, and backward steps are used to compute the unobserved state vector and error covariance matrices. The training error is the difference between observation values and PMU measurement data. 
As for the prediction/online anomaly detection step, the trained, dynamic regression model is used to predict the frequency measurement values for the next 5 seconds. Since the sliding window is 5 minutes with 1 second resolution and the prediction is for 5 seconds, this is a short-term prediction, and it can capture the anomalies when the prediction deviates by 3.5 standard deviations from the training error.

\section{Results and Discussion}
This section discusses the experimental results and corresponding performance data of PMU 5G communication, along with the signal delay and quality analysis. The unique IP and port were assigned for each node for this experiment. 

\subsection{5G PMU-PDC Communication}
\label{subsec:pdc}
\begin{figure}[!ht]
\centering
\includegraphics[width=.85\linewidth]{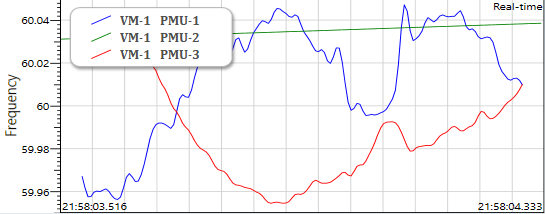}
\caption{PMU and PDC communication through 5G network and the C37.118 data were received at openPDC.}
\label{fig:openpdc}
\end{figure}

Following the configuration of 5G PMU communication, we proceeded to simulate the process of transmitting real-time streaming PMU data from the sensor to the data concentrator through the 5G network. The real-time C37.118 data was generated by the PMU streaming data generator within the PlayPDAT ESXI \textbf{VM-1}. Fig. \ref{fig:openpdc} shows the results of the open PDC at streaming data receiver VMs. The chart in openPDC displayed three sample signals that were collected from 3 real PMU devices. This test examined that 5G technology can be applied for PMU to PDC wireless communication following IEEE C37.118 standard.

\subsection{5G Network Communication Performance Analysis}
To perform the communication network analysis, Wireshark was utilized to identify the timestamp at both the transmitter and receiver end, evaluating the transmission latency of C37.118 streaming data over the 5G network. The Wireshark can track the packet details, such as time, IP, destination, protocol, packet size, etc. By checking the sending and receiving timestamps from two Wireshark installed at PlayPDAT VM and openPDC VM, along with the PTP time synchronization, we can roughly estimate the data transmission latency by the difference value of two recorded timestamps.  

\begin{figure} [!hbt]
    \centering
    \includegraphics[width=.9\columnwidth]{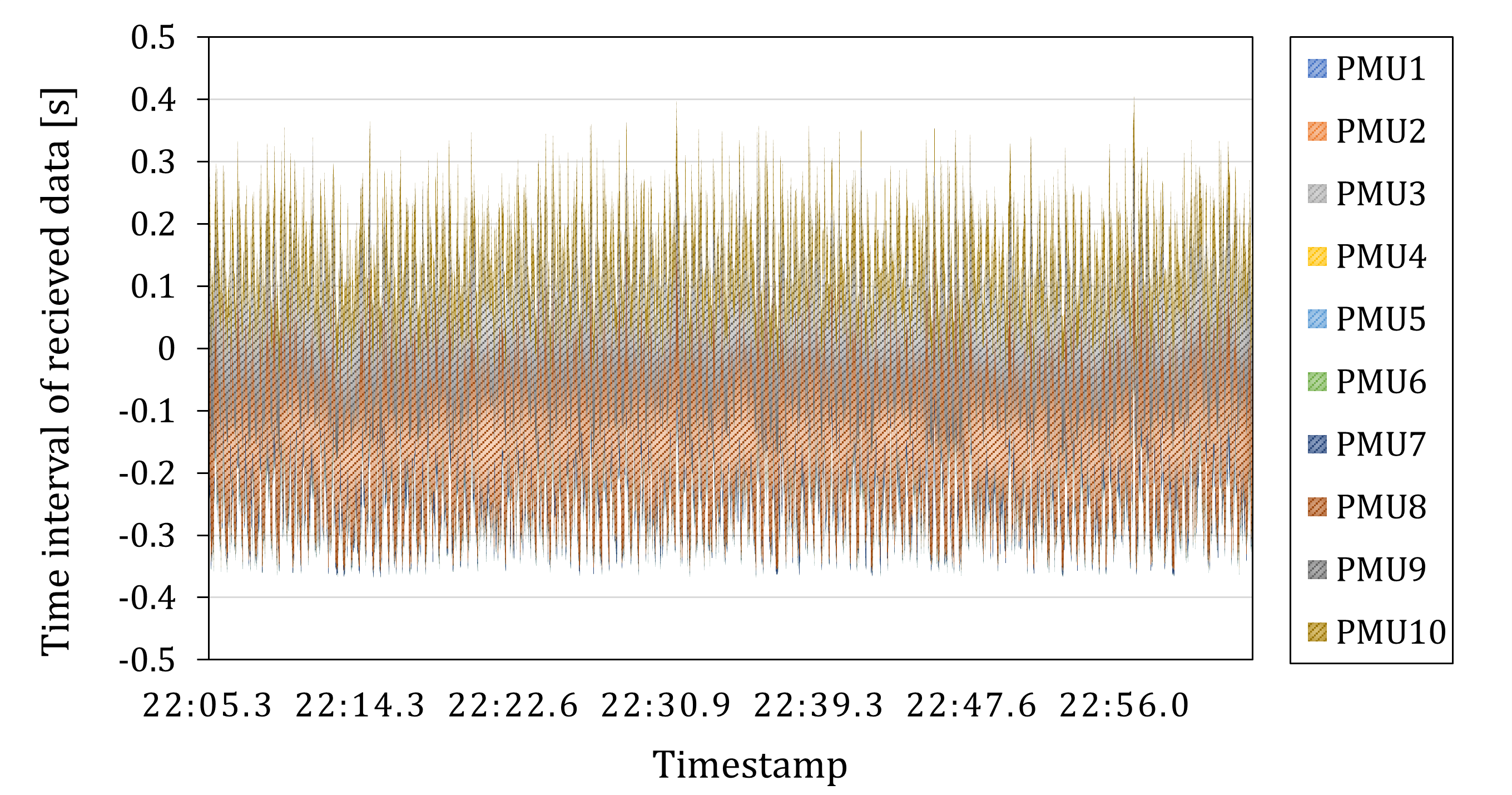}
    \caption{Trace plot for PMU communication over 5G network.}
    \label{fig:10trace}
\end{figure}

To ensure the experimental reliability and generalization, we collected 10 trace trials of PMU-PDC data from the testing environment. Visual representations of these traces were generated by plotting the packet number against the time gap between PlayPDAT and openPDC, illustrated in Fig. \ref{fig:10trace}. Each trial encompassed 1 minute of recording, s.t., 60 seconds of 60 FPS PMU data. 

Interestingly, the time gap observations showed that some trials had a significant discrepancy from others, i.e., the average time gap for trial 4 is more than 0.5s, but others are not. By counting the packet number sent from PlayPDAT and received at openPDC , we observed that the packets received at receiving (openPDC) end were not exactly 60$\times$60 number of packets, which caused packet duplication and loss in UDP. 

As there is no identifier information in UDP for detecting duplicate packets, the data from a packet may be indicated twice (or even more) to the application. After filtering out the duplication and loss packets, the delay was estimated. The average delay was close to 9.063ms, which was in compliance with the delay requirement of the C37.118.2011 Annex C2.

\subsection{ML-Application Results}

\begin{figure}[!ht]
    \centering
    \includegraphics[width=.85\columnwidth]{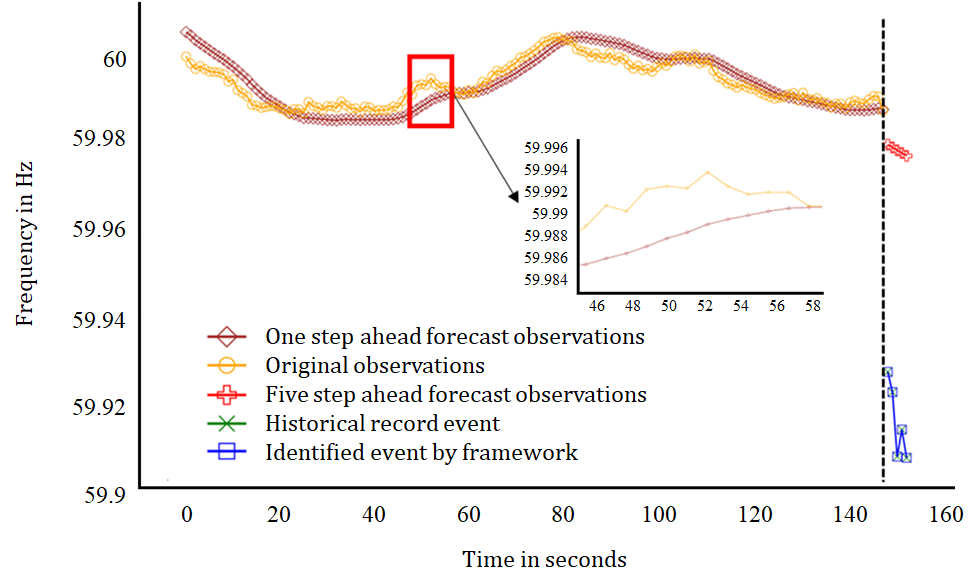}
    \caption{Detection of anomalous event at edge device using the PMU data stream via 5G network testbed.}
    \label{fig:figmain}
\end{figure}

As described in Section~\ref{subsubsec:prediction}, a 5-minute sliding window with a 1-second refreshing rate was implemented for real-time anomaly detection using the GDLM model. Fig.~\ref{fig:figmain} showcases the edge anomaly detection for the PMU data stream received via the 5G network test bed. As shown in the inset of Fig.~\ref{fig:figmain} (red box), it can be observed that prediction (curve fitting results on the sliding window data) are very close to the original observations differing only by the third decimal place in the frequency attribute. This shows that the GDLM anomaly detection model provides satisfactory goodness of fit during the model fitting stage (left-hand side of the black dotted line in Fig.~\ref{fig:figmain}). 

Furthermore, when the event occurs at the $148^{\text{th}}$ second (vertical black dashed line), it can be observed that the 5-step ahead prediction for the next 5 seconds has a significant discrepancy between previous measurements and predicted values. This difference between the prediction and the measurements proves that the GDLM model can successfully capture anomalous streaming data in the test PMU data as discussed in Section~\ref{subsubsec:prediction}, the 3.5 standard deviation is used as the threshold to detect the anomalies successfully. This helps GDLM distinguish the prediction error and actual anomalies in a moving window. Therefore, the results show that GDLM anomaly detection can be deployed at 5G edge for protection or corrective control action to prevent potential system cascading failures \cite{9494800}.

\section{Conclusion}
In this paper, we have successfully integrated 5G telecommunication technology into PMU-PDC communication, employing the UDP transmission protocol. Subsequently, we conducted a comprehensive analysis of the quality of the 5G network using collected traces. In addition, an ML-based anomaly detection application deployed on 5G-enabled edge computing is demonstrated. The test outcome reveals the feasibility of deploying the 5G in PMU communication based on C37.118, enabling PMU-PDC to exchange substantial real-time measurements efficiently. Besides, with 5G edge computation, anomaly detection can be achieved with extremely low latency, improving power system monitoring and analytics. The 5G PMU communication was realized solely on UDP for time-sensitive commitment. Due to its board-casting functionality, some potential packet loss might occur. In the future, we will test other communication protocols, such as TCP and other industrial protocols. Also, we intend to utilize the feature inherent to 5G technology to realize distribution power system protection for inverter-based renewable energy resources.

\bibliographystyle{IEEEtran}
\bibliography{references}

\begin{thebibliography}{10}
\providecommand{\url}[1]{#1}
\csname url@samestyle\endcsname
\providecommand{\newblock}{\relax}
\providecommand{\bibinfo}[2]{#2}
\providecommand{\BIBentrySTDinterwordspacing}{\spaceskip=0pt\relax}
\providecommand{\BIBentryALTinterwordstretchfactor}{4}
\providecommand{\BIBentryALTinterwordspacing}{\spaceskip=\fontdimen2\font plus
\BIBentryALTinterwordstretchfactor\fontdimen3\font minus \fontdimen4\font\relax}
\providecommand{\BIBforeignlanguage}[2]{{%
\expandafter\ifx\csname l@#1\endcsname\relax
\typeout{** WARNING: IEEEtran.bst: No hyphenation pattern has been}%
\typeout{** loaded for the language `#1'. Using the pattern for}%
\typeout{** the default language instead.}%
\else
\language=\csname l@#1\endcsname
\fi
#2}}
\providecommand{\BIBdecl}{\relax}
\BIBdecl

\bibitem{5535240}
F.~Li, W.~Qiao, H.~Sun, H.~Wan, J.~Wang, Y.~Xia, Z.~Xu, and P.~Zhang, ``Smart transmission grid: Vision and framework,'' \emph{IEEE Transactions on Smart Grid}, vol.~1, no.~2, pp. 168--177, 2010.

\bibitem{5524055}
P.~Zhang, F.~Li, and N.~Bhatt, ``Next-generation monitoring, analysis, and control for the future smart control center,'' \emph{IEEE Transactions on Smart Grid}, vol.~1, no.~2, pp. 186--192, 2010.

\bibitem{9905351}
O.~Muhayimana and P.~Toman, ``A review on phasor measurement units and their applications in active distribution networks,'' in \emph{2022 IEEE PES/IAS PowerAfrica}, 2022, pp. 1--5.

\bibitem{9917615}
N.~Anton, C.~Bulac, M.~Sănduleac, E.-E. Gemil, B.~Dobrin, and V.-A. Ion, ``An overview of pmu-based electrical power systems modelling for power quality enhancement,'' in \emph{2022 57th International Universities Power Engineering Conference (UPEC)}, 2022, pp. 1--4.

\bibitem{9787983}
S.~Bu, L.~G. Meegahapola, D.~P. Wadduwage, and A.~M. Foley, ``Stability and dynamics of active distribution networks (adns) with d-pmu technology: A review,'' \emph{IEEE Transactions on Power Systems}, vol.~38, no.~3, pp. 2791--2804, 2023.

\bibitem{IEC61850}
\emph{Communication networks and systems for power utility automation}, International Electrotechnical Commission Standard IEC 61\,850, 2020.

\bibitem{6083199}
A.~A. Zambrano, M.~A. Leon, and E.~Rivas, ``Phasor measurement unit using gprs wireless connectivity,'' in \emph{2011 IEEE PES CONFERENCE ON INNOVATIVE SMART GRID TECHNOLOGIES LATIN AMERICA (ISGT LA)}, 2011, pp. 1--7.

\bibitem{6469217}
D.~Ghosh, T.~Ghose, and D.~K. Mohanta, ``Communication feasibility analysis for smart grid with phasor measurement units,'' \emph{IEEE Transactions on Industrial Informatics}, vol.~9, no.~3, pp. 1486--1496, 2013.

\bibitem{9797119}
M.~Maksimovic, M.~Forcan, M.~C. Boškovic, T.~B. Šekara, and B.~Lutovac, ``On the role of 5g ultra-reliable low-latency communications (urllc) in applications extending smart grid (sg) capabilities,'' in \emph{2022 11th Mediterranean Conference on Embedded Computing (MECO)}, 2022, pp. 1--4.

\bibitem{943067}
``{IEEE Standard for Synchrophasers for Power Systems},'' \emph{IEEE Std 1344-1995(R2001)}, pp. i--, 1995.

\bibitem{ieeestd}
``{IEEE Standard for Synchrophasor Data Transfer for Power Systems},'' \emph{{IEEE Std C37.118.2-2011 (Revision of IEEE Std C37.118-2005)}}, pp. 1--53, 2011.

\bibitem{6804630}
``{IEEE Standard for Synchrophasor Measurements for Power Systems -- Amendment 1: Modification of Selected Performance Requirements},'' \emph{IEEE Std C37.118.1a-2014 (Amendment to IEEE Std C37.118.1-2011)}, pp. 1--25, 2014.

\bibitem{openpdc}
``{Open Source Phasor Data Concentrator},'' \url{https://github.com/GridProtectionAlliance/openPDC}, accessed: 2023-03-08.

\bibitem{wireshark}
``{Wireshark},'' \url{https://www.wireshark.org/}, accessed: 2023-03-08.

\bibitem{4579760}
``{IEEE Standard for a Precision Clock Synchronization Protocol for Networked Measurement and Control Systems},'' \emph{IEEE Std 1588-2008 (Revision of IEEE Std 1588-2002)}, pp. 1--269, 2008.

\bibitem{faris2016bpa}
T.~Faris, ``Bpa synchrophasor lab tools,'' in \emph{WECC JSIS meeting, September}, 2016.

\bibitem{ren2018online}
H.~Ren, Z.~Hou, and P.~Etingov, ``Online anomaly detection using machine learning and hpc for power system synchrophasor measurements,'' in \emph{2018 IEEE International Conference on Probabilistic Methods Applied to Power Systems (PMAPS)}.\hskip 1em plus 0.5em minus 0.4em\relax IEEE, 2018, pp. 1--5.

\bibitem{9494800}
C.~Hannon, D.~Deka, D.~Jin, M.~Vuffray, and A.~Y. Lokhov, ``Real-time anomaly detection and classification in streaming pmu data,'' in \emph{2021 IEEE Madrid PowerTech}, 2021, pp. 1--6.

\end{thebibliography}
\end{document}